# To What Extent Does Not Wearing Shoes Affect the Local Dynamic Stability of Walking? Effect Size and Intrasession Repeatability

Philippe Terrier,[1,2] Fabienne Reynard[1,2]

[1]IRR, Institute for Research in Rehabilitation, Sion, Switzerland
[2]Clinique romande de réadaptation SUVACare, Sion, Switzerland

**Funding:** The study was not supported by specific funding. The Institute for Research in Rehabilitation (IRR) is financially supported by the State of Valais, the City of Sion, the Clinique Romande de Réadaptation, and the Swiss accident insurance company SUVA.

**Conflict of Interest Disclosure:** No known competing interests.

**Correspondence Address:**

Dr Philippe Terrier
Clinique romande de réadaptation SUVACare
Av. Gd-Champsec 90
1951 Sion
Switzerland
Tel.: +41-27-603-23-91
E-mail: Philippe.Terrier@crr-suva.ch

**Running title:** Gait stability during barefoot walking

**Abstract**

Local dynamic stability (LDS) quantifies how a system responds to small perturbations. Several experimental and clinical findings have highlighted the association between gait LDS and fall risk. Walking without shoes is known to slightly modify gait parameters. Barefoot walking (BW) may cause unusual sensory feedback to individuals accustomed to shod walking (SW), and this may impact on LDS. The objective of this study was therefore to compare the LDS of SW and BW in healthy individuals and to analyze the intrasession repeatability. Forty participants traversed a 70 m indoor corridor wearing normal shoes in one trial and walking barefoot in a second trial. Trunk accelerations were recorded with a 3D-accelerometer attached to the lower back. The LDS was computed using the finite-time maximal Lyapunov exponent method. Absolute agreement between the forward and backward paths was estimated with the intraclass correlation coefficient (ICC). BW did not significantly modify the LDS as compared to SW (average standardized effect size: +0.12). The intrasession repeatability was high in SW (ICC: 0.73–0.79) and slightly higher in BW (ICC: 0.82–0.88). Therefore, it seems that BW can be used to evaluate LDS without introducing bias as compared to SW, and with a sufficient reliability.

**Introduction**

Although bipedal locomotion is inherently unstable,[1] human beings exhibit a high resilience to external perturbations that could lead to falls.[2, 3] However, many pathologies and conditions may alter the capabilities to efficiently manage obstacles and perturbations while walking. In particular, fall-related injuries are a major health issue in elderly people.[4, 5] Analytical methods derived from nonlinear analysis of dynamical systems have been proposed to analyze gait stability and related fall-risks.[6] Following the nomenclature on stability theory and nonlinear dynamical systems, the largest perturbation that individuals can tolerate without falling is referred to as "global stability"[7] or "stability margins."[8] Inside those margins, motor control continuously adjusts gait parameters to compensate for infinitesimal perturbations induced, for instance, by neuromuscular noise. This is referred to as "local dynamic stability" (LDS).[9] LDS can be characterized using the maximal Lyapunov exponent, which is a parameter that assesses how infinitesimal perturbations grow over time (butterfly effect) or, in other words, how fast the system diverges.[6, 9-11] Rosenstein et al.[12] proposed a practical method for calculating maximal Lyapunov exponents from small data sets. With this method, local divergence exponents ($\lambda$) are computed from the slopes of divergence curves, which quantify how fast the neighboring trajectories of a reconstructed state space diverge from nearest neighbor points ("initial perturbation"). Strictly speaking, due to the nonlinearity of the divergence curves, multiple slopes could be defined. Hence, no true single maximal Lyapunov exponent exists. Different slopes (divergence exponents) quantify local divergence (and hence local stability) at different time scales. Classically, two different time scales have been proposed to assess gait LDS: long-term divergence $\lambda_l$[6] (long-term LDS) and short-term divergence $\lambda_s$[7] (short-term

LDS). The former is based on the time interval between four and 10 strides after the initial perturbation, and the latter is based over a time interval corresponding to one stride (or one step) after the initial perturbation. Authors first proposed to use long-term LDS to assess gait stability.[6] However, recent developments in the field have shown that short-term LDS is a more responsive index.[13, 14] Consequently, throughout the article, the generic LDS term is used to refer to short-term LDS, and the time scale reference is used when necessary.

Although there are no theoretical reasons why local stability should predict global stability,[13] it is assumed that if motor control can efficiently manage small perturbations (low divergence inside the stability margins), it can also thwart large perturbations that would lead to falling. Hence, studies have suggested that LDS might predict global stability and fall risk.[6, 7, 9] Recent theoretical[13, 14] and experimental[15, 16] results have supported this hypothesis. For instance, 3D gait modeling has shown that LDS is responsive to noise added to the model and that it serves as an early predictor of fall risk.[13] Furthermore, recent clinical studies have shown that elderly subjects at risk for falling exhibited lower LDS.[17] Based on the aforementioned recent fundamental and clinical researches, it is increasingly clear that LDS may be a valid fall-risk indicator.

The recent development of LDS as a clinically valid index for the follow-up of various pathologies has attracted growing interest.[18] However, there is still a need for further studies to translate the fundamental research results into an operational clinical tool. In particular, potential bias and confounding factors that could inadvertently modify LDS data should be thoroughly documented. For instance, studies have shown that LDS could be influenced by the length of the measurement,[19, 20] turning during the walking test,[21] the use of a treadmill,[7, 11] or the walking speed.[22]

Using ice-induced plantar desensitization, Manor et al.[23] observed a substantial reduction in gait short-term LDS (-40%), with only a minor change in long-term LDS (-8%). Furthermore, it has been observed that the nature of the walking surface (compliance) could modify long-term LDS but not short-term LDS.[24] This suggests that tactile sensations at the level of the foot, as well as the proprioception and position of the foot, could have a different impact on short-term and long-term LDS. Furthermore, studies have been reported that barefoot walking (BW) induces slightly shorter steps and higher walking cadence.[25-27] Such a modification of gait pattern could be an indication of a more cautious gait, which has been associated with an increased fall risk in older people.[28] It could therefore be assumed that individuals who wear shoes most of the time would experience some difficulties in optimally managing gait stability when walking barefoot due to unusual sensory feedback from the feet. Despite this potential confounding factor, LDS studies have not systematically reported the footwear status of their participants. Many studies seem to have measured shod walking (SW),[29, 30] whereas others have evaluated BW.[31, 32] It is unclear whether LDS results obtained in shod individuals can be compared to those obtained in barefoot individuals.

To use LDS as a gait quality index suitable for individual assessments in a clinical context, it is crucial to evaluate the absolute agreement of LDS between consecutive measurements performed in the same individual, either using a short-term perspective (intrasession reliability) or a longer-term perspective (intersession reliability). For clinical applications, it is of paramount importance that repeatability results define the minimal detection threshold in the LDS change at the individual level. The intrasession repeatability of treadmill walking,[19] as well as the intra- and interday reliability of outdoor walking,[33] has been already studied. However, there is still a need to better characterize the reliability of LDS in short indoor walking tests that could be used in clinical settings.

The main objective of the present study was to compare the LDS of SW and BW in healthy middle-aged individuals. Short-term and long-term LDS were analyzed because previous studies have shown that both parameters may be relevant.[23, 24] The hypothesis was that BW would induce lower LDS because the absence of shoes might produce unusual sensory feedback to individuals accustomed to SW. As a secondary goal, the study aimed to assess the intrasession repeatability (absolute agreement between measurements) in order to enhance the generalizability of the BW vs. SW results and to provide reference values applicable to short walking tests.

## Methods

### Subjects

Forty healthy individuals [19 males, 21 females; mean (*SD*): 37 years (10), with height 1.72 m (0.08) and body mass 68 kg (13)] participated in the study. All the subjects gave their written informed consent. The study was approved by the regional medical ethics committee (Commission Cantonale Valaisanne d'Ethique Médicale, Sion, Switzerland).

### Procedure

The participants were instructed to walk straight ahead at a self-selected comfortable walking speed along a 70 m hallway and then to do a U-turn and return. The walking surface was hard, corresponding to standard hospital flooring. The participants wore their own shoes. They were instructed to wear shoes in which they felt quite comfortable, but no high heels were allowed. They performed one trial (2 ×70 m) with shoes and one trial barefoot. The sequence between the BW and the SW trials was randomized. Trunk accelerations were recorded with a tri-axial accelerometer (Physilog system, BioAGM, Switzerland), which was attached to the lower back (over the spine, L3-L4 level) with a belt and connected to a lightweight data logger (Physilog system, BioAGM, Switzerland; sampling rate 200 Hz, 16-bit resolution). The accelerometer measured the body accelerations along three axes: medio-lateral (ML), vertical (V), and antero-posterior (AP). Subsequent data analysis was performed with Matlab (Mathworks, MA, USA). Statistical analysis was realized in part with measures of effect size (MES) toolbox.[34]

**Data Analysis**

Following graphical inspection, U-turns were discarded from the raw acceleration signals, and two segments of steady gait were selected (one for the forward path and one for the backward path). The step frequency (SF, Hz) was computed by fast Fourier transform of the raw acceleration signals. Eighty steps were then selected. To improve the normalization of the data, the 3D acceleration signals pertaining to the 80 steps were time-normalized at 8,000 samples (resampling), ensuring a constant sample length.

The method for quantifying the LDS from divergence exponents has been described in detail in numerous articles.[6, 7, 11, 35] Interested readers will find a thorough theoretical background in the review by Dingwell.[9] Here, only the parameters necessary to reproduce the results are summarized. The state space was reconstructed according to Takens' theorem, as classically applied in gait dynamics studies.[6] The embedding dimension and the time delay were assessed by global false nearest neighbors (GFNN) analysis and the average mutual information (AMI) function, respectively. A constant dimension of six was set for all the directions. A time delay of 10, 12, and 12 samples, respectively, was used for the ML, V, and AP directions. These values corresponded to the average results of the GFNN and AMI analyses. The maximum finite-time Lyapunov exponents ($\lambda$) were estimated from the slopes of the linear fits in logarithmic divergence diagrams, as defined by Rosenstein's algorithm.[12] Time was normalized by the average stride time (1/SF) in each trial, taking into account the resampling. The divergence exponents were computed over a time scale corresponding to one step (0.5 stride, short-term LDS $\lambda_s$)[7, 13] and over the fourth to the 10th strides (long-term LDS $\lambda_l$).[6]

**Statistics**

To analyze the footwear effect, the results of both the forward and backward paths were averaged together. Notched boxplots (median and quartiles) were used to describe the data (Fig. 1 and 2). The means and standard deviations (*SD*) are also presented, including the average change and the corresponding *SD* (BW minus SW). The coefficient of variation (CV = *SD*/mean × 100) was used to assess interindividual variability.

The standardized effect size (ES) is reported using Hedges's *g*,[36] which is a variant method of Cohen's *d* for inferential measures. The difference between SW and BW was the contrast measure, and the standardizer was the pooled standard deviation ($s_p$). Ninety-five percent CIs were estimated by bootstrapping.[34] The results are presented (Fig. 3) with the arbitrary limits for small (0.2), medium (0.5) and large (0.8) ES.[37] To minimize type I error risk induced by the numerous comparisons, the analysis was completed with a multivariate comparison test (Hotelling's T-squared).

The null hypothesis (H0) was that the mean differences (BW-SW) along the three axes were equal to zero.

To explore whether BW induced a kinematics change that would explain the change in LDS, the correlation (Perason's *r*) between the change in cadence (SF) and the change in LDS (BW minus SW) was assessed, with corresponding 95% CIs (asymptotic estimates).

Both the intraclass correlation coeffecient (ICC) and the standard error of measurement (SEM) were used to characterize the intrasession reliability.[38] Both forward and backward paths were separately analyzed as two intrasubject repetitions. The approach was that proposed by McGraw and Wong,[39] based on the classical work of Shrout and Fleiss.[40] The ICC(1) model was used, which assesses the degree of absolute agreement among measurements made on randomly selected objects (one-way model).[39] This approach was justified by the fact that the study was focused on the intrasession reliability evaluated in two trials, which were consecutively measured with no changes in the measurement method. The agreement between the two repetitions was separately analyzed under the two conditions (SW and BW). The 95% CIs on the ICC were computed using traditional *F* statistics.[39] SEM is the group-level estimation of the intrasubject average variability (expected trial-to-trial noise in the data). It was computed with the following equation: $SEM = S_T \sqrt{1-R}$, where $S_T$ is the grand *SD*, (i.e., the *SD* of the pooled data of the two repetitions), and *R* is the ICC result. To facilitate the comparison among the parameters, the CV (i.e., SEM/grand mean × 100) was also computed.

Finally, the Spearman–Brown prophecy formula was used to predict the number of strides necessary to achieve high reliability (i.e. $R_A^* = 0.90$), taking into account that a normalized number of 40 strides was tested for repeatability. The formula was as follows:

$$N = \frac{R_A^*(1-R_A)}{R_A(1-R_A^*)} \qquad (5)$$

where $N$ is the estimated number of trials needed to achieve the expected level of repeatability $R_A^*$ given the observed repeatability $R_A$. The number of trials $N$ was converted to the corresponding number of strides. An example of this approach applied to the field of gait analysis can be found in the work of Hollman and others.[41]

## Results

Regarding the cadence results, there was a small increase in BW condition compared to the SW condition (Fig. 1). In other words, individuals tended to walk at a higher step rate when they walked barefoot. The absolute effect corresponded to 2.4 steps·min$^{-1}$ (+2%). Compared to the average *SD* among the individuals (0.13), the *SD* of the difference (0.04) was small, indicating a high homogeneity of the response to BW among the participants (relative agreement between BW and SW, Pearson's *r* = 0.92). The interindividual variability (CV) was 4.3% (SW) and 5.7% (BW).

Regarding LDS (Fig. 2), the results revealed relative changes, which ranged from 7% to 11% for $\lambda_l$ and from -0.6% to +5% for $\lambda_s$ (lower $\lambda$ indicates higher LDS). For $\lambda_l$, the interindividual variability (CV) was (SW) 33%, 27%, and 27%, respectively, and (BW) 37%, 24%, and 28%, respectively, in the ML, V, and AP directions. For $\lambda_s$, the interindividual variability (CV) was (SW) 12%, 13%, and 15%, respectively, and (BW) 16%, 15% and 17% in the ML, V, and AP directions.

Regarding the ES results (Fig. 3), a small (ES: +0.34) but significant effect of BW was confirmed for cadence. Long-term LDS ($\lambda_l$) exhibited an average ES of - 0.36, implying that BW is more locally stable, with a barely significant effect in the AP and ML directions. However, the multivariate comparison indicated that there was no significant effect when the three axes were compared together (Hotelling's $T^2$ = 6.85, $p$ = 0.11). No substantial changes were observed in the short-term LDS ($\lambda_s$) in the ML and AP directions (mean ES = -0.01), but a small effect (decreased stability, ES = 0.33) was evident along the vertical axis. According to the result of the multivariate test (Hotelling's $T^2$ = 8.66, $p$ = 0.06), there appeared to be no significant overall difference between SW and BW.

The correlation analyses revealed that no relevant relationship existed between cadence changes (ΔSF) and LDS changes (Δλ) (BW minus SW). The results were ($r$ and 95% CI, $N$ = 40): ΔSF vs. $\Delta\lambda_l$ ML, $r$ = 0.08 (-0.24−0.38); ΔSF vs. $\Delta\lambda_l$ V, $r$ = -0.01 (-0.31−0.30); ΔSF vs. $\Delta\lambda_l$ AP, $r$ = 0.01 (-0.30−0.32); ΔSF vs. $\Delta\lambda_s$ ML, $r$ = 0.16 (-0.15−0.45); ΔSF vs. $\Delta\lambda_s$ V, $r$ = -0.14 (-0.44−0.17); ΔSF vs. $\Delta\lambda_s$ AP, $r$ = -0.00 (-0.32−0.31).

Regarding reliability results, a high repeatability was present under both conditions regarding cadence (SF). On the contrary, poor repeatability was observed for long-term LDS, with the ICC ranging from 0.22 to 0.63. There were also high within-subject errors (SEM CV: 23–29%). The repeatability was higher for short-term LDS, with the ICC ranging from 0.74 to 0.87 and the SEM CV ranging from 6% to 8%. BW walking induced a more consistent LDS compared to SW.

The analysis of the ICC results using the Spearman–Brown prophecy formula (Table 1) confirmed the low reliability of long-term LDS (211–1276 strides to reach 90% reliability) and the sufficient reliability of short-term LDS (54–126 strides).

## Discussion

In 40 healthy individuals, by analyzing 3D trunk accelerations during short walking trials, this study aimed to analyze the difference between BW and SW in terms of LDS. No relevant effects were observed. Consequently, the hypothesis that postulated an effect of BW on gait stability mediated through the modification of sensory feedback should be discarded. Furthermore, the repeatability results showed that short-term LDS can be assessed with high reliability but that long-term LDS exhibits poor reliability.

As in most recent LDS studies,[15, 42, 43] this study employed a normalized number of strides (40) and a normalized sample size (8,000). It also used standardized parameters for state space reconstruction (uniform dimension [6] and time delays [10, 12, 12]). Finally, as proposed by others,[15, 43] this study computed short-term LDS over one step, and not one stride. Exploratory analysis of preliminary data (not shown) revealed that those choices yielded the highest repeatability. Many studies of gait LDS used a treadmill to obtain standardized experimental conditions.[7, 11, 22, 44] A treadmill makes it possible to impose a large range of walking speeds other than the preferred walking speed. By imposing substantial changes in walking speed, it has been demonstrated that speed has an influence on LDS.[22, 45, 46] On the other hand, testing overground walking allows physiological walking conditions to be analyzed. In overground, unconstrained walking conditions, it has been shown that individuals exhibit very low stride-to-stride variability (CV <3%) in their gait parameters, including their preferred walking speed,[47] due to energetic optimization of locomotion.[48] The high resilience of motor control prevents gait parameters, such as speed or cadence, diverging from optimal values. In particular, studies have observed that BW has quite a limited effect on preferred walking speed.[26, 27] As a result, caution should be exercised when extrapolating treadmill results to overground situations. A reanalysis of the results obtained by Bruijn et al.[22] can serve as an illustration of the problem. In a treadmill experiment, they explored the influence of a large range of speeds (0.62 to 1.72 m·s$^{-1}$) on LDS with a similar method as in the present study. The difference between 1.28 m·s$^{-1}$ and 1.06 m·s$^{-1}$ (-20% relative change) can be used to roughly extrapolate what would induce a 5% decrease in speed: short-term LDS, AP -1%, ML 0%, V +2%; long-term LDS, AP +4%, ML -6%, V +6%. Regarding the low responsiveness of LDS to changes in speed under the range of physiological variations, the possibility of a potential change in speed between BW and SW having a relevant effect can be excluded with a high confidence.

As the present study included a substantial number of individuals ($N = 40$) with a large range of ages (18–58 years), the results are very likely generalizable to a healthy adult population. Furthermore, the ad-hoc reliability results facilitate an assessment of the effect of measurement errors and intraindividual variability. By averaging the results of the two trials together (i.e., 80 strides analyzed), the expected reliability for short-term LDS is between 85% and 93%. Thus, about 90% of the total variance was due to actual between-subject variance. However, it is worth noting that the participants were European people accustomed to SW, with low experience of BW. In addition, the study measured only acute effects. The effects after a longer habituation time to BW may be different.

The increase of cadence in BW is a well-documented phenomenon, especially in children.[25] In adults, it has been observed that when they are not wearing shoes, they tend to walk with shorter steps and at a higher step rate.[26] The same study reported that cadence increased by +2.8%, the step length decreased by -5.0%, and hence the speed decreased by -2.3%. Another study reported the following effects in a comparison of BW with SW (boots): cadence +5.4%, step length -6.4%, speed, -1.4%.[27] Accordingly, the results of the present study confirmed a small increase in SF (+2.4 steps·min$^{-1}$, +2%) in BW. The correlation results demonstrated that this change was not related to a change in the LDS. In other words, individuals that exhibited a greater change in cadence did not exhibit a concomitant change in LDS. All the correlation coefficients were below 0.2. The extent of the CIs shows that correlations higher than 0.4 are very unlikely at the population level.

The following experimental results can help to place the results in a broader context. By applying visual and mechanical perturbation to walking individuals, Sinitksi et al.[44] found that LDS was reduced by about -25%. By impairing balance control by randomly varying galvanic vestibular stimulation (GVS), van Schooten et al.[15] reported an effect on LDS of -11%. In another study, treadmill walking significantly increased LDS compared to overground walking by +9%.[11] The effect of aging, defined as the relative difference between young and older adults, on LDS has been found to be about -70%[49] or -40%.[50]

Although omnibus testing ($T^2$, $p = 0.11$) revealed no significant modification of long-term LDS, it should be taken into account that the low observed repeatability (Fig. 4) greatly increased the risk of type II statistical error, which is also highlighted by the large CIs (Fig. 3). Therefore, the existence of an effect at the population level cannot be excluded. Namely, the average ES (Fig. 3, -0.36), the significant changes along the ML and AP directions, and the average relative change (Fig. 2, -11%), which was in the range of the reported change in the literature appear to suggest that individuals exhibit more local stability in BW. Many studies have shown that long-term LDS is poorly related to actual fall risk.[13, 14, 17] However, other authors have suggested that enhanced long-term LDS could be related to compensatory mechanisms under destabilizing situations[15] or to a more cautious gait.[24] Those factors may constitute a valid explanation for the results in the present study.

Unlike long-term LDS, short-term LDS exhibited high repeatability. Thus, a lower risk of type II error is expected. Short-term LDS exhibited no significant change in multivariate testing, but the results obtained were close to the 5% significance level ($T^2$, $p = 0.06$). This is mainly due to the small (ES: 0.32, relative change: +5%) but significant destabilizing effect (higher $\lambda_s$) that was observed along the vertical axis. As compared to the results of the other studies (see above),[11, 15, 44] the effect is probably of limited relevance. Conversely, no change in the ML and the AP directions (average ES: 0.03) were observed. The spread of the CIs excludes with high confidence that a substantial effect exists at the population level.

As step duration is a highly controlled parameter, which exhibits low stride-to-stride variability even during long duration walking,[47] the very high observed repeatability in SF (0.96, Fig. 4) is not surprising. Similar values have been described in the literature.[26, 41, 51] To our knowledge, only two studies have been dedicated to the assessment of LDS repeatability.[19, 33] By using treadmill walking, Kang and Dingwell compared different walking durations (1–5 min) during three repetitions. They reported that at least 3 min were necessary to reach good repeatability (ICC >0.75) for short-term LDS, whereas long-term LDS ICC leveled-off around 0.6. They observed that in 1 min walking, the ICC was around 0.45 for short-term LDS and around 0.30 for long-term LDS. Recently, using a similar method to that employed in the present study, van Schooten et al.[33] analyzed both intra- and intersession repeatability during long-duration outdoor walking (500 m). Only short-term LDS was analyzed. They found that the ICC was around 0.8 for intrasession repeatability. Although the current study used shorter walking tests, we found higher repeatability than Kang's study and similar results as in van Schooten's study. On the other hand, we confirmed that long-term LDS exhibited large intraindividual variability (CV SEM 23–30%), which severely compromises its use at the individual level.

From the recent literature, it has become increasingly clear that short-term rather than long-term LDS is the most appropriate parameter to assess global stability and fall risk.[16, 17, 52] Furthermore, the importance of lateral LDS has been emphasized.[16] The present study showed that this parameter was not modified in BW. Therefore, healthy individuals seem able to maintain optimal dynamic stability, even when faced with the unusual situation of walking without shoes, despite the fact that they probably adopt a slightly more cautious gait.[26] Furthermore, high intrasession repeatability was observed in the present study. Consequently, short duration walking tests might be appropriate to assess gait stability, even to measure differences between conditions at the individual level. Fifty-four strides could be sufficient to reach 90% reliability (Table 1). In addition, performing BW tests seems to further enhance repeatability. Furthermore, in longitudinal studies or in comparisons between groups, it is not evident to standardize footwear because individuals wear varying type of shoes. Thus, analyzing BW instead of SW may improve the standardization in LDS assessment.

**Figure captions**

**Figure 1** Descriptive statistics of the step frequency (SF) of shod (SW) and barefoot (BW) walking

Boxplots show quartiles, the median, and the spread of the data across study group ($N = 40$). Values are means (*SD*). Bold values are the average change (BW minus SW) and the corresponding *SD*.

**Figure 2** Descriptive statistics of gait stability of shod (SW) and barefoot (BW) walking

Results of both short-term local dynamic stability (LDS) (bottom) and long-term LDS (top) are shown. Boxplots show quartiles, the median, and the spread of the data across study group ($N = 40$). The values are the means and the standard deviations (*SD*). Bold values are the average change (BW minus SW) and the corresponding *SD*. ML, medio-lateral; V, vertical; AP, antero-posterior.

**Figure 3** Effect size of barefoot walking (BW) compared to shod walking (SW)

The small filled circles show the standardized mean difference (Hedges's *g*), and horizontal lines are the corresponding 95% confidence intervals. Negative (positive) values indicate that BW induced a lower (higher) value of the observed variable compared to shod walking. SF, step frequency. LDS, local dynamic stability. ML, medio-lateral; V, vertical; AP, antero-posterior.

**Figure 4** Intrasession repeatability

The intrasession repeatability was estimated using the intraclass correlation coefficient (ICC(1), two repetitions and 40 subjects). The ICC values are printed on the left. The small black circles are the graphical representation of the ICCs, with the corresponding 95% confidence intervals. Continuous lines are the results for shod walking (SW), and dashed lines are the results for barefoot walking (BW). The standard errors of measurement (SEM) are shown on the right, with the corresponding coefficient of variation (CV). SF, step frequency; LDS, local dynamic stability; ML, medio-lateral; V, vertical, AP, antero-posterior.

Author's preliminary draft. Journal of Applied Biomechanics 2013*Author's preliminary draft. Journal of Applied Biomechanics 2013*

**Table**

**Table 1** Prediction of the number of strides necessary to reach 90% repeatability

|  |  | Number of strides | |
|---|---|---|---|
|  |  | SW | BW |
| Cadence | SF | 15 | 19 |
| Long-term LDS | $\lambda_l$-ML | 240 | 211 |
|  | $\lambda_l$-V | 699 | 518 |
|  | $\lambda_l$-AP | 1276 | 640 |
| Short-term LDS | $\lambda_s$-ML | 126 | 54 |
|  | $\lambda_s$-V | 84 | 59 |
|  | $\lambda_s$-AP | 114 | 84 |

The Spearman–Brown prophecy formula was used with the ICC presented in Figure 4 as an input. SW, shod walking; BW, barefoot walking; SF, step frequency; LDS, local dynamic stability; ML, medio-lateral; V, vertical; AP, antero-posterior.

Figure 1

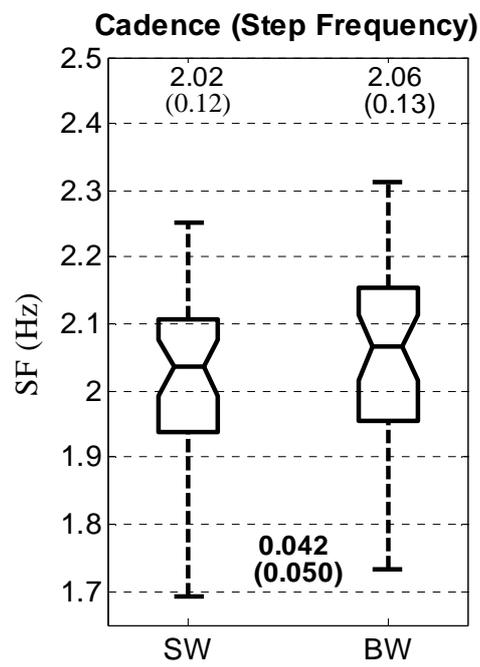

Figure 2

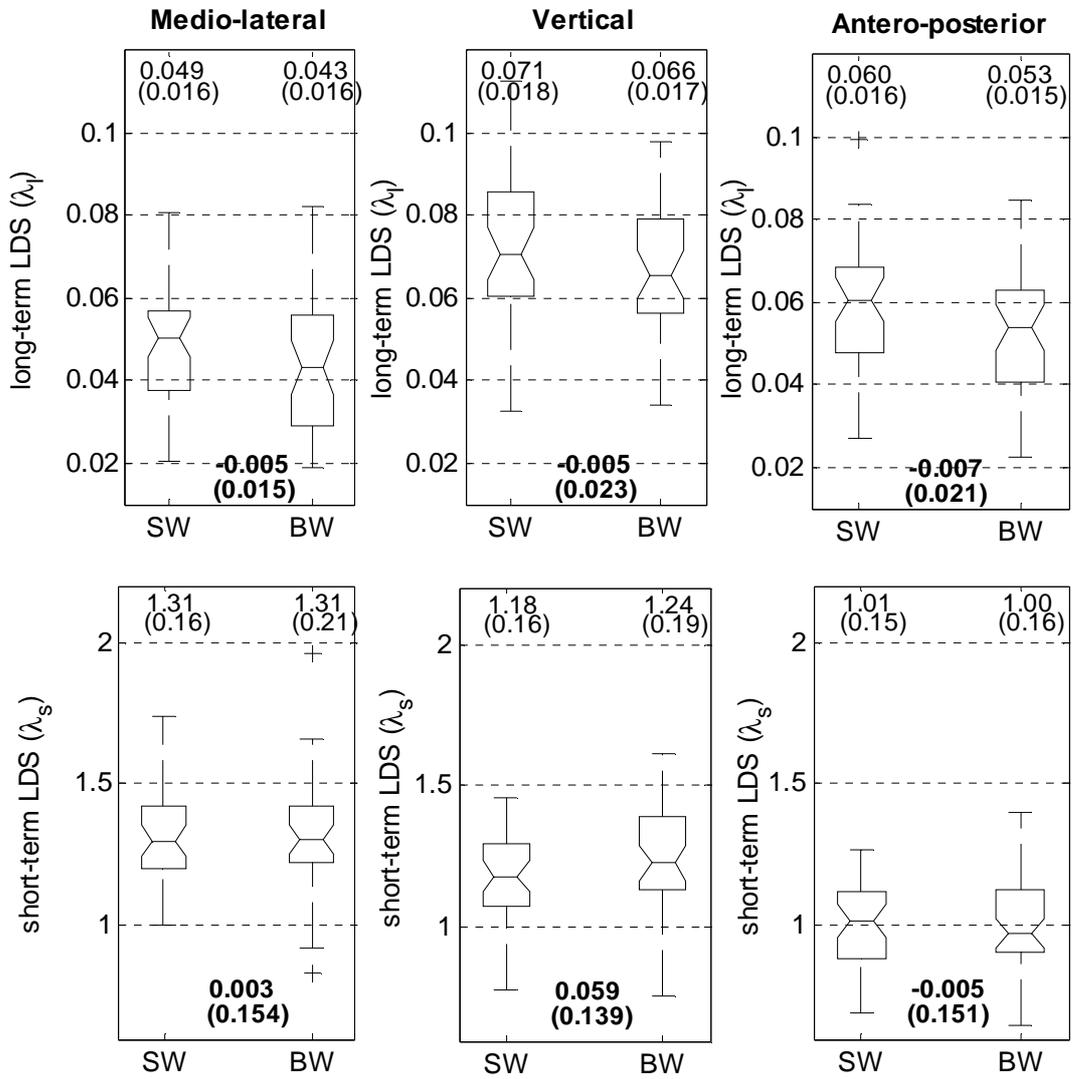

Figure 3

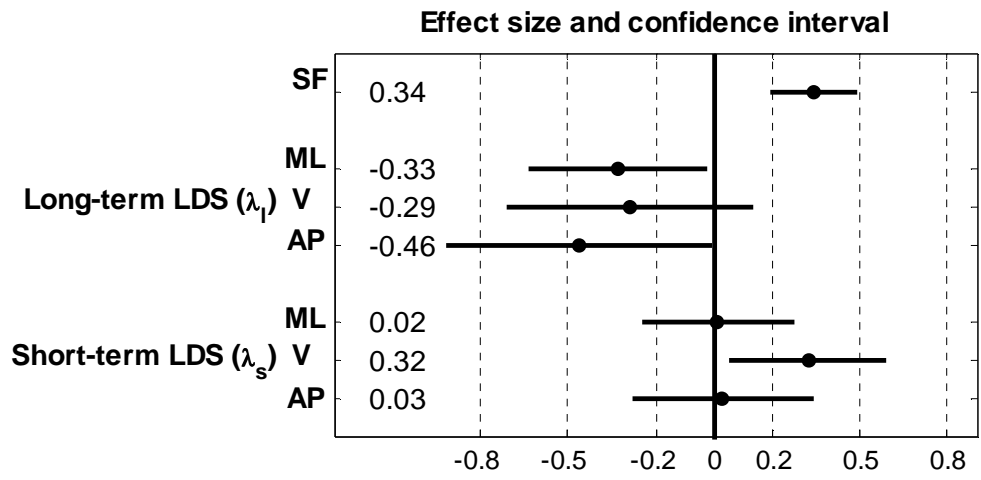

Figure 4

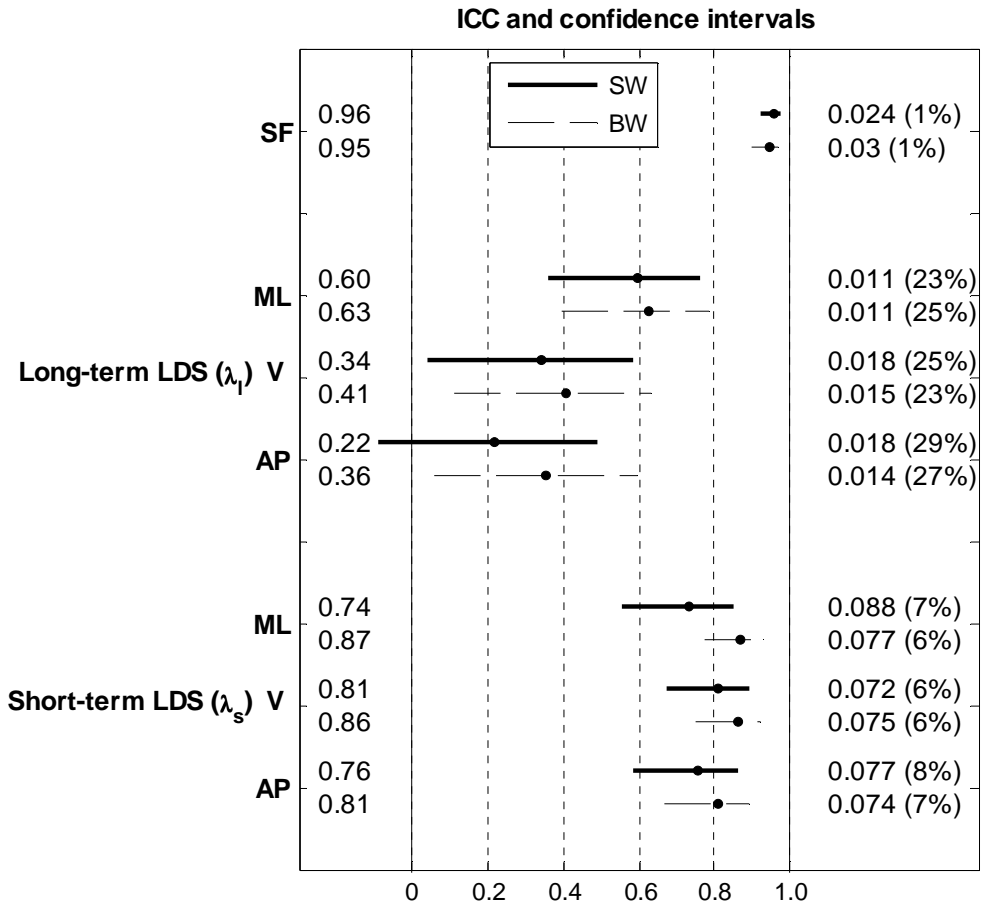